# Bright betatron radiation from direct-laser-accelerated electrons at moderate relativistic laser intensity

O. N. Rosmej[1,2,3], X. F. Shen[4], A. Pukhov[4], L. Antonelli[5], F. Barbato[6], M. Gyrdymov[2], M. M. Günther[1], S. Zähter[1], V. S. Popov[7,8], N. G. Borisenko[9], N. E. Andreev[7,8]

[1]GSI Helmholtzzentrum für Schwerionenforschung GmbH, Planckstr.1, 64291 Darmstadt, Germany
[2]Goethe University, Frankfurt, Max-von-Laue-Str. 1, 60438 Frankfurt am Main, Germany
[3]Helmholtz Forschungsakademie Hessen für FAIR (HFHF), Campus Frankfurt am Main, Max-von-Laue-Straße 12, 60438 Frankfurt am Main, Germany
[4]Heinrich-Heine-University Düsseldorf, Universitätsstraße 1, Düsseldorf, Germany
[5]York Plasma Institute, University of York, Church lane, Heslington, York, YO10 5DQ, U.K.
[6] University of Bordeaux, CNRS, CEA, CELIA, UMR 5107, F-33405, Talence, France
[7] Joint Institute for High Temperatures, RAS, Izhorskaya st.13, Bldg. 2, 125412 Moscow, Russia
[8] Moscow Institute of Physics and Technology (State University), Institutskiy Pereulok 9, 141700 Dolgoprudny Moscow Region, Russia
[9] P. N. Lebedev Physical Institute, RAS, Leninsky Prospekt 53, 119991 Moscow, Russia

o.rosmej@gsi.de, GSI Helmholtzzentrum für Schwerionenforschung GmbH, Planckstr.1, 64291 Darmstadt, Germany

ABSTRACT

Direct laser acceleration (DLA) of electrons in a plasma of near critical electron density (NCD) and associated synchrotron-like radiation are discussed for moderate relativistic laser intensity (the normalized laser amplitude $a_0 \leq 4.3$) and ps-long pulse. This regime is typical for kJ PW-class laser facilities designed for high energy density research. Currently, in experiments at the PHELX laser it was demonstrated that interaction of $10^{19}$ W/cm$^2$ sub-ps laser pulse with sub-mm long NCD plasma results in generation of high-current well-directed super-ponderomotive electrons with effective temperature that is 10x higher than the ponderomotive potential [O. Rosmej *et al*., PPCF **62,** 115024 (2020)]. Three-dimensional Particle-In-Cell simulations provided a good agreement with the measured electron energy distribution and were used in the current work to study synchrotron radiation of the DLA accelerated electrons. The resulting x-ray spectrum with a critical energy of 5 keV reveals an ultra-high photon number of $7\times10^{11}$ in the 1-30 keV photon energy range at the focused laser energy of 20 J. Numerical simulations of a betatron x-ray phase-contrast imaging based on the DLA process for the parameters of a PHELIX laser is presented. The results are of interest for applications in high energy density (HED) experiments, which require a picosecond x-ray pulse and a high photon flux.

## I. INTRODUCTION

Synchrotron-like radiation generated by electrons during laser-plasma acceleration is characterized by a broadband x-ray spectrum with a small divergence and a high brilliance. It provides an excellent tool for hard x-ray radiography and x-ray phase contrast high-resolution imaging in biology[1-3], medicine[4], and in investigations of ultra-fast processes in laser-driven shock waves[5,6] or implosion[7,8]. A detailed overview of the betatron radiation applications is given in[9]. The characteristics of betatron radiation such as the critical energy $E_c$, photon number, bandwidth, source size, and divergence angle depend on the mechanism of the electron acceleration, electron energy distribution and charge carried by electrons.

Extensive experimental studies of the synchrotron radiation were performed at sub-100 fs pulse laser systems in the LWFA-regime of electron acceleration[10-12] where the production of quasi-mono-energetic GeV electron beams was demonstrated[13,14] in under-dense plasma with electron density of ~ $10^{18}$ cm$^{-3}$. Electrons confined in the focusing phase of the plasma wake experience transverse betatron oscillations, emitting highly collimated synchrotron x-rays. In such experiments, the charge carried by electrons was far below the nC-level and the number of photons around the critical energy $E_c$ of some keV does not exceed $10^9$. Nevertheless, a short pulse duration and low divergence angle resulted in a record brilliance of $10^{23}$ photons/s/mm$^2$/mrad$^2$/0.1%BW reported in[12].

Increase of the plasma density and/or laser pulse duration at relativistic laser intensity can lead to a longitudinal breakup ("self-modulation") of the laser pulse[15,16]. For the self-modulated regime of acceleration (SMLWFA), the laser pulse must be substantially longer than the period of the Langmuir electron oscillations. As a result, the wake takes a form of a rapidly evolving train of electron cavities driven by relativistic intense optical bullets. This turns the electron spectrum into quasi-Maxwellian, while increasing the accelerated charge by more than one order of magnitude compared to the LWFA-case. Measurements of the betatron radiation, generated in the self-modulated LWFA regime at the sub-ps sub-kJ Titan laser system, are reported in[17,18]. Up to $10^9$ photons/ (eV sr) at 6 keV were measured in the interaction of a 150 J laser pulse with an under-dense plasma of ≤ $10^{19}$ cm$^{-3}$ electron density. Simulations performed for the 1.1 kJ sub-ps PETAL-conditions [8] showed that SMLWFA regime predicts up to $10^{12}$ emitted photons in the energy range of 2-60 keV and a high brilliance of $5\times10^{20}$ photons/s/mm$^2$/mrad$^2$/0.1%BW.

Simulations of synchrotron radiation generated in NCD plasma by direct laser accelerated electrons were performed for ultra-relativistic laser intensities ≥ $10^{21}$ W/cm$^2$ and fs-short laser pulses[19,20]. These simulations demonstrated ≥ $10^{10}$ MeV-photons in 0.1% BW with critical energy of 5 MeV as a result of electron acceleration up to 1 GeV energy and high transverse focusing force at high plasma density of $n_e$ =1.8 $n_{cr}$[20] (the critical electron density is defined as $n_{cr} = m_e \omega_L^2 /(4\pi e^2)$ where $m_e$ and $e$ are the mass of electron at rest and its charge and $\omega_L$ is the laser frequency).

In this work, we explore synchrotron radiation in long-scale NCD plasma at moderate relativistic laser intensity, for parameters typical for kJ PW-class laser facilities currently in operation, and demonstrate a feasibility of a DLA-based betatron source that promises an ultra-high number of x-ray photons for applications in HED research. The use of low density foams has opened the door to the practical realization of DLA in NCD plasma and has demonstrated its high robustness.

In pilot experiments performed at PHELIX [21,22], long-scale NCD plasma with a thickness of 300-500 µm and ≤ $10^{21}$ cm$^{-3}$ electron density was produced using a well-controlled ns-pulse that triggers a supersonic ionization wave[23] in low density polymer aerogels[24]. Accelerated electrons with effective temperature of 10-20 MeV and electron energies up to 100 MeV carry up to µC charge[21,22]. 3D PIC simulations performed for the PHELIX-laser intensity of $2.5\times10^{19}$ W/cm$^2$ and 20 J contained in the FWHM of the focal spot result in the synchrotron radiation with $7\times10^{11}$ photons (1-30 keV) and a critical energy of $E_c \sim 5$ keV.

The paper is organized as follows. In Section II, experimental data[22], which demonstrate realization of the DLA process in sub-mm long pre-ionized polymer foams are presented. In Section III, parameters of the betatron radiation obtained in 3D PIC simulations based on the experimental results achieved on the PHELIX facility are discussed. The DLA-based betatron source is compared with synchrotron radiation generated in the LWFA and SMLWFA regimes at different laser systems. Section IV presents simulations of the x-ray phase contrast imaging for radiographic applications in HED experiments. Section V summarizes the results.

## II. DLA IN PLASMA OF NEAR CRITICAL DENSITY

The electron acceleration in relativistic laser channels generated by intense laser pulse in NCD plasma was discussed for the first time in [25,26]. The authors proposed a mechanism of direct laser energy coupling into hot electrons that requires strong self-generated static electric and magnetic fields (Fig.1). Relativistic laser pulse propagating in the NCD plasma experiences strong self-focusing. Ponderomotive expulsion of background plasma electrons from the plasma channel creates a radial electrostatic field $E_r$ that has a pinching polarity for electrons. At the same time, the current of accelerated electrons generates a strong ($\geq 100$ MG)[21,23] azimuthal magnetic field $B_\phi$ that traps electrons in the ion channel. The betatron radiation is generated when the relativistic electrons undergo transverse betatron oscillations in these self-generated quasi-static fields. An efficient electron energy gain occurs when the betatron frequency becomes close to the Doppler shifted laser frequency[25, 26].

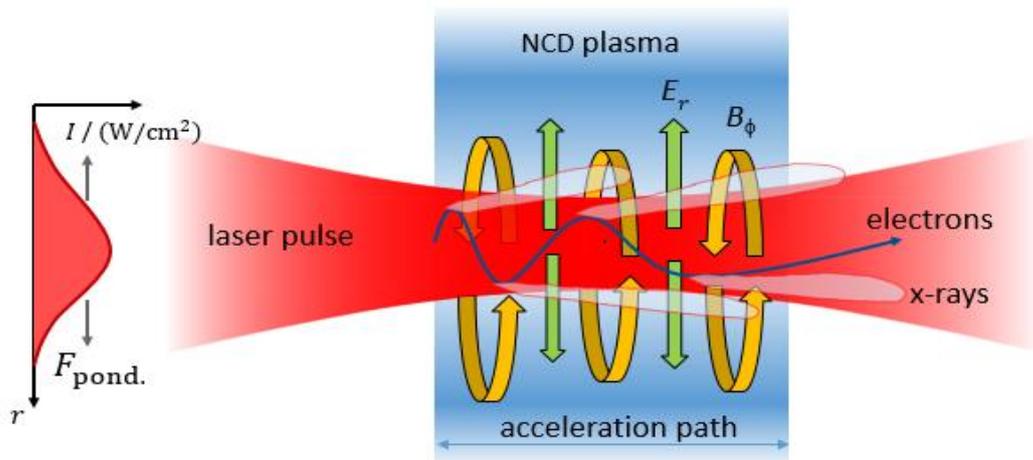

**FIG. 1.** Betatron radiation is generated when relativistic electrons undergo transverse betatron oscillations in self-generated quasi-static electric and magnetic fields.

The DLA works efficiently in plasmas of near critical density (NCD) and in the case of sub-ps laser pulses. To explore this acceleration mechanism, pilot experiments [21,22] were performed on the Petawatt High Energy Laser for Ion eXperiments (PHELIX) at the GSI Helmholtzzentrum Darmstadt[28]. A s-polarized laser pulse of 1.053 µm fundamental wavelength and energy $E_{las} \simeq$ 80-100 J delivered by the Nd:glass laser was focused into an elliptical focal spot with FWHM diameters (12 ± 2) µm and (18 ± 2) µm using a 150 cm off-axis parabolic mirror. Laser energy in the FWHM of the focal spot was $E_{FWHM} \simeq$ (17-22) J [22]. The peak laser intensity reached (1-2.5) ×10$^{19}$ W/cm$^2$ ($a_0$ = 2.7- 4.28) at 750 ± 250 fs laser pulse duration. Here $a_0$ is the normalized vector potential that scales as $a_0^2 = 0.73 \cdot I_{L,18} \cdot \lambda^2$ with the laser intensity $I_{L,18}$ normalized to $10^{18}$ W/cm$^2$ and the laser wavelength $\lambda$, in µm. Triacetate cellulose layers of 2 mg/cm$^3$ volume density and 300 - 400 µm thickness[26] were used as targets. A sub-mm long NCD plasma was produced by sending a well-defined ns pulse forerunning the relativistic main pulse onto a foam. The intensity of the ns laser pulse was kept at ~ 5x10$^{13}$ W/cm level in order to initiate a super-sonic ionization wave propagating with 2×10$^7$ cm/s velocity, for more details see[21, 23].

In experiment[22], electron spectra were measured in one shot by means of 0.99 T static magnets and imaging plates (IP) used as detectors. Spectrometers were placed at 0°, 15° and 45° to the laser axis in the plane perpendicular to the laser polarization at 40 cm from the interaction point. Figure 2(a) presents electron spectra measured by irradiation of a pre-ionized 2 mg/cm$^3$ CHO-foam of 325 µm thickness at ~ 2×10$^{19}$ W/cm$^2$ laser intensity. The electron energy distributions show a strong angular dependence with a highest effective temperature and highest electron number close to the laser axis. The electron spectra were approximated by Maxwellian-like distribution functions with one or two temperatures: at 0° with T = (12.0 ±1.4 MeV), at 15° with $T_1 \simeq$ 8.0 MeV ($T_2 \simeq$ 11.0 MeV), and at 45° with T $\simeq$ 2.9 MeV. The angular distribution of electrons with energies > 3 MeV in a wide range of angles was measured by means of a stack of three stainless steel cylindrical plates[22]. It was shown that super-ponderomotive electrons propagate in a divergence cone of 0.16 sr.

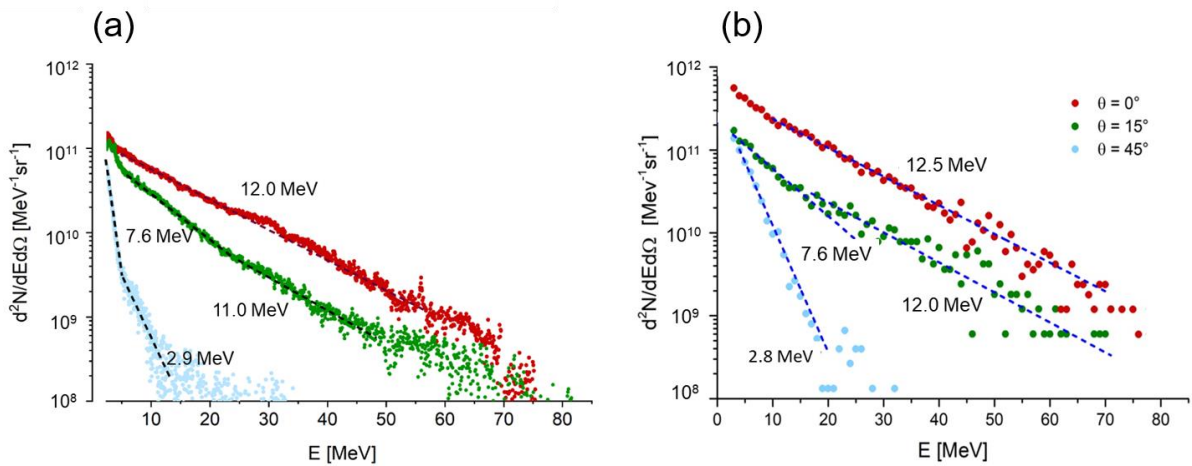

**FIG. 2.** (a) Energy distribution of super-ponderomotive electrons per steradian measured at 0°(red), 15°(green) and 45°(blue) to the laser propagation direction for a shot onto the pre-ionized foam layer at 2×10$^{19}$ Wcm$^{-2}$ laser intensity, (b) PIC-simulations for the same interaction parameters [O. Rosmej *et al*., PPCF **62,** 115024 (2020)].

The simulated spectra shown in Fig. 2 (b) describe very well the measured electron energy distributions for different angles. The number of simulated super-ponderomotive electrons is up to 2-5 times higher than measured in the experiment. The experimental number can be corrected upwards by at least factor 2-3 due to the loss of particles by propagation through 20 mm long WCu collimators placed in front of magnetic spectrometers. One suggests that the reason for this effect is a quasi-static electric field created at the collimator front by x-rays or/and high current relativistic electrons that reached the collimator at the very beginning of the interaction[29]. Electrons that approach the collimator later can be slightly deviated by this field from their initial trajectory and cannot pass through the narrow entrance slit. One can suppose that this effect is even stronger for electrons detected at 0° to the laser axis due to their higher number. This problem will be investigated in more details in forthcoming experiments.

## III. PIC-SIMULATIONS OF BETATRON RADIATION

3D PIC simulations of the DLA-process in NCD plasma and associated synchrotron-like radiation were performed for the laser parameters and interaction geometry described in[24] using the Virtual Laser Plasma Laboratory (VLPL) code[30]. Laser pulse intensity in time and space was approximated by the Gaussian distribution with elliptical form of the focal spot taken from the experiment with FWHM axes 11 μm in vertical and 15 μm horizontal direction. The laser pulse energy in the FWHM focal spot of 20 J and the FWHM pulse length of 700 fs resulted in the laser intensity of $2.5\times10^{19}$ W/cm$^2$ with $a_0 = 4.28$. The homogeneous plasma was composed of electrons and fully ionized ions of carbon, hydrogen and oxygen. Simulations accounted for the ion type and the ion fraction in accordance with the chemical composition of triacetate cellulose $C_{12}H_{16}O_8$.

Betatron radiation is produced when relativistic electrons undergo transverse betatron oscillations in the self-generated magnetic and electrostatic fields. The fundamental dimension-less parameter of betatron radiation is $K = \gamma k_\beta r_\beta$, where $\gamma$ is the relativistic factor of electrons, $k_\beta$ and $r_\beta$ are the betatron wave number and the amplitude of electron betatron oscillations, respectively. In the case of $K \gg 1$ (wiggler regime), the radiation spectrum is quasi-continuous broadband and has a shape given by the universal function $S(\omega/\omega_c)$[31] where $S(x) = x \int_x^\infty K_{5/3}(\xi)d\xi$, $K_{5/3}$ is the modified Bessel function of the second kind and $\omega_c = 3K\gamma^2 c k_\beta / 2$ is the critical frequency, where $c$ is the speed of light. For frequencies well below the critical frequency, the function $S(x)$ increases with frequency as $\omega^{2/3}$, reaches a maximum at $\approx 0.29\omega_c$, and then drops exponentially to zero above $\omega_c$. For discussed experimental conditions[22], a wiggler strength reaches K ~ 60 and the critical photon energy may be expressed as

$$E_c = \hbar\omega_c = \frac{3}{2}\hbar\gamma^3 c r_\beta k_\beta^2 \qquad (1)$$

For the electron beam, which is characterized by the effective electron temperature > 10 MeV and charge of ~μC, one expects an intense betatron radiation with a large photon number generated in the ion channel during the DLA process. In VLPL, at every time step, we assume that the relativistic electrons emit photons in the direction of its propagation with the spectrum defined by $S(\omega/\omega_c)$, where $\omega_c$ is determined by $\omega_c = (3/2)\gamma^2 |F_\perp|/(m_e c)$ and $F_\perp$ is the transverse force felt by electrons[31,32]. The radiation recoil effect is also included into the electron motion equations. This approach has been widely verified by experiments and independent simulations[10,12,33,34].

Figure 3(a) shows the spectral distribution of the betatron radiation simulated for the PHELIX laser parameters ($a_0 = 4.28$, 0.7 ps, $E_{FWHM} = 20$ J) presented in semi-logarithmic scale. The betatron spectrum is peaked at 1.7 keV and has the critical energy of 5 keV. The x-ray photon number reaches $6\times10^{11}$ in 1-10 keV and $1\times10^{11}$ in 10-60 keV energy ranges. A 0.1% BW (band width) at ~ 5 keV contains ~ $3\times10^8$ photons. According to simulations, the laser pulse in NCD plasma undergoes strong self-focusing resulting into 4 µm (root mean square, RMS) transverse radius of the radiation source. Source opening angles at the critical energy in $Y$ (polarization) and Z-directions are rather large and reach $\theta_{FWHM}$: 840 mrad and 640 mrad correspondingly. In spite of the large divergence, mentioned above characteristics of the DLA-based betatron source ensure a high brilliance $B(\theta_{FWHM}) \simeq 6\times10^{19}$ ph/s/mm$^2$/mrad$^2$/0.1% BW.

Fig. 3(b) shows a 2D map of the photon fluence at the detector at a distance of 120 cm from the source. The distribution of the radiation is not very homogenous, one can observe two hot spots in polarization direction, which are 2-3 times more intense than the surrounding area. The photon fluence at the detector remains at a very high level, up to $10^8$ x-ray photons/cm$^2$, which is very promising for applications in HED-research.

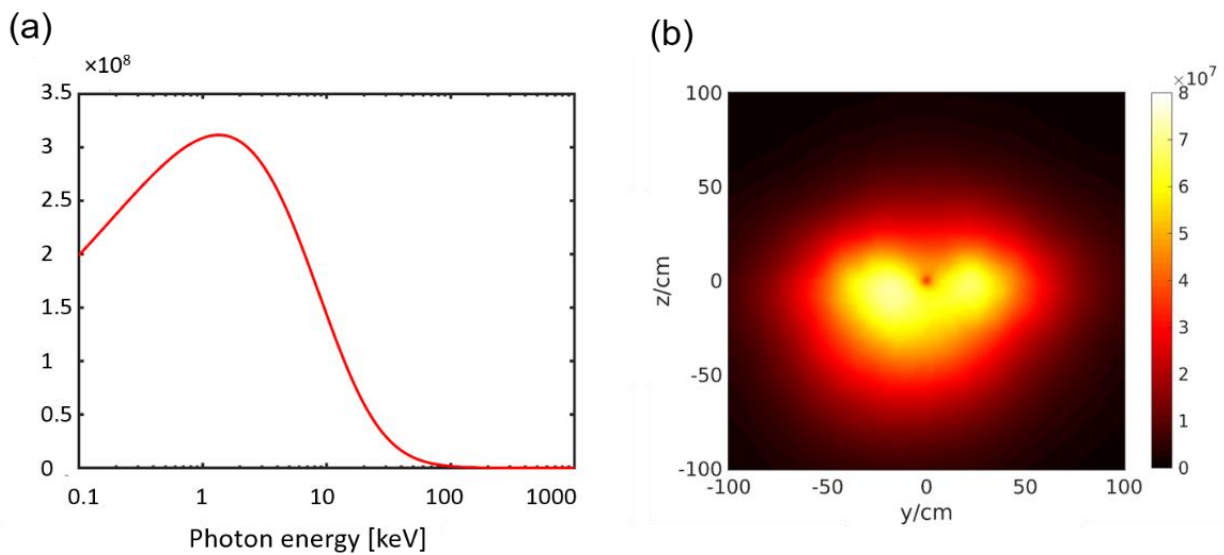

**FIG. 3.** (a) Spectral distribution of the betatron radiation simulated for the PHELIX parameters, (b) 2D map of the photon fluence at the detector placed at a distance of 120 cm from the source.

Table 1 presents characteristics of the betatron-sources produced via different mechanisms of the electron acceleration such as LWFA, SMLWFA and DLA. In the table, parameters of the laser-target interaction and characteristics of generated ultra-relativistic electrons and betatron-sources are shown. The betatron radiation produced in the LWFA-process has an ultra-high brilliance caused by a small source-size, a low divergence and tens of fs pulse duration[12]. On the other hand, a moderate number of photons is expected due to the relatively small charge carried by electrons accelerated in the under-dense plasma.

With increasing the laser pulse duration and/or plasma density, a transition from LWFA to a SMLWFA takes place[15,16]. Experiment[17] and simulations[8] demonstrate an increase of the electron bunch charge up to 10-50 nC and high x-ray photon fluence. Since the laser pulse undergoes self-focusing by propagation in under-dense plasma, the x-ray source size is smaller than the initial laser focus. Nevertheless, the brilliance suffers due to a higher x-ray beam divergence and longer pulse duration than in the LWFA-case (Table 1).

In the DLA-process, μC charge of super-ponderomotive is generated in NCD plasma[21,22]. An ultra-high x-ray photon number of ~ $7\times10^{11}$ is predicted by numerical simulations performed for laser and target parameters used in the experiment ($E_{las}$ ~ 80 J ($E_{FWHM}$ ~ 20 J), 0.7 ps, $a_0 = 4.2$; target: 300 µm long NCD layer of $6.5\times10^{20}$ cm$^{-3}$ electron density). This photon number is similar to that obtained for the SWLWFA-based betatron source in simulations performed for the PETAL laser parameters (1.1 kJ, 0.5 ps, $a0 = 7$) and centimetres long plasma target of $(1-3)\times10^{18}$ cm$^{-3}$ electron density (see Table 1, Ferii *et al.*). The reason for this is more than two orders of magnitude higher plasma density and subsequently a high number of DLA-accelerated electrons. High background plasma density is also responsible for a strong transverse focusing force $F_\perp$ acting on the relativistic electrons in the plasma channel. Since $\omega_c = (3/2)\gamma^2 |F_\perp|/(m_e c)$ [31,32], the photon critical energy in the DLA-case differs by a factor of only 2-5 from LWFA and SM-LWFA, despite significantly lower Lorentz factor γ of accelerated electrons. Drawback of the high electron density (high focusing force) is a 10 times larger than in the SMLWFA-case[8] opening angle.

| Laser/Target | Regime | Electrons | x-rays | Brilliance | Reference |
|---|---|---|---|---|---|
| **ASTRA-GEMINI** 50 fs, 5J, 35 µm $a_0 \simeq 2$ $N_e \simeq 10^{18}$ cm$^{-3}$ | Resonant betatron oscillations in plasma wake experiment | $E_{max} \simeq 700$ MeV $Q_e$ N/A | $N_{ph} \simeq 5\times10^8$ $E_c \simeq 50$-450 keV $\theta_{FWHM} = 14$ mrad Source size $L_{RMS} \simeq 15$ µm | $10^{23}$ | S. Cipiccia *et al.*, Nature Phys. **7** (2011) |
| **TITAN** 0.7 ps, 120 J, 20 µm $a_0 \simeq 3$ $N_e \simeq 10^{19}$ cm$^{-3}$ supersonic gas-jet | Self-Modulated LWFA experiment | $T_{hot} \simeq 15$ MeV $E_{max} \simeq 300$ MeV $Q_e \simeq 10$ nC | $N_{ph} \simeq 10^9$/eV sr $\Delta E_{ph} \simeq 6.5\pm0.5$ keV $E_c \simeq 10$ keV $\theta_{FWHM} = $ ---------- $L_{RMS} \simeq 35$ µm | N/A | F. Albert *et al.*, PRL *118* (2017), Nucl.Fus.59 (2018) |
| **PETAL** 0.5ps, 1 kJ, 42 µm $a_0 \simeq 7.5$ $N_e \simeq (1-3)\times 10^{18}$ cm$^{-3}$ cms long plasma | Self-Modulated LWFA CALDER-CIRC simulations | $E_e \simeq 1$ GeV $Q_e \simeq 38$ nC (>70 MeV) | $N_{ph} \simeq 7\times10^{11}$ $\Delta E_{ph} \simeq 2$-60 keV $E_c \simeq 10$ keV $\theta_{FWHM} = 50$ mrad $L_{RMS} \simeq 25$ µm | $5\times10^{20}$ | J. Ferii *et al.*, Phys. Rev. Accel. beams **19** (2016) |

| PHELIX<br>0.7 ps, 80 J ($E_{FWHM} \simeq$ 20-30 J), 15 µm<br>$a_0 \simeq 3 - 4$<br>$N_e \simeq 0.6 \times 10^{21}$ cm$^{-3}$<br>pre-ionized low density polymer aerogels | DLA at betatron resonance<br><br>experiment<br>3D PIC | $T_{hot} \simeq 13$ MeV<br>$E_{max} \simeq 100$ MeV<br><br>$Q_e \simeq 1$ µC<br>(> 2 MeV)<br><br>$Q_e \simeq 100$ nC<br>(> 7 MeV)<br><br>experiment | $N_{ph} \simeq 6 \times 10^{11}$<br>$\Delta E_{ph} \simeq$ 1-10 keV<br><br>$N_{ph} \simeq 10^{11}$(>10keV)<br>$E_c \simeq 5$ keV<br><br>$\theta_{FWHM} \simeq 700$ mrad<br>$L_{RMS} \simeq 4$ µm<br><br>3D PIC | $6 \times 10^{19}$<br>($\theta_{FWHM}$) | O. Rosmej et al., PPCF **57** (2020)<br><br>Current work |
|---|---|---|---|---|---|

Table 1. Comparison of betatron sources produced in LWFA, SMLWFA and DLA-processes.

Comparing performance of LWFA and DLA-driven betatron sources, it is also important to consider their application fields. The LWFA-based betatron radiation provided by sub-100 TW short pulse lasers was successfully applied to radiograph biological objects[1-3]. The photon number of $10^8$-$10^9$ was sufficient for XPCI in one shot. In the HED-research[18,36,37], kJ PW- class lasers produced a high level of parasitic background caused by accelerated particles and radiation interacting with target chamber and diagnostic set-up. In these hazard conditions, an ultra-high photon number is required in order to ensure a high signal-to-noise ratio of XPCI-technique.

Scaling the DLA-process towards higher laser energy keeping laser pulse duration and focal spot size constant will result in a grow of the laser field amplitude and a consequent increase of the electron energies and number of the DLA-electrons. In NCD plasma at high laser power, the acceleration process can be hindered by the filamentation and hosing instabilities of the laser pulse propagating in plasma channel. 3D PIC simulations performed for higher laser energy and intensity demonstrated that also in this case the DLA process continues to be very effective[27, 35] and can ensure enhanced production of the betatron radiation.

## IV. X-RAY PHASE CONTRAST IMAGING SIMULATIONS FOR PHELIX

High energy PW-class laser systems such as PHELIX are designed for HED research and the study of matter in extreme states. In this context, time-resolved x-ray absorption and x-ray phase contrast imaging (XPCI) radiography using bremsstrahlung emission was performed on the PHELIX to study a laser-driven shock-wave travelling in a target material[36, 37]. The nanosecond evolution of the shock-wave requires an x-ray pulse in the picoseconds regime and high photon flux to freeze the phenomenon in a single-shot experiment. Bremsstrahlung sources can be very bright, however, the isotropic emission can produce a large noise induced by the irradiation of the entire chamber and components. In this sense, the betatron radiation represents an improvement because of its directionality. Moreover, betatron is generated in low Z polymer foams, while bremsstrahlung sources are obtained irradiating high Z target to increase the photon yield. The directionality of the betatron x-ray beam allows to place both, object and detector at larger distance from the x-ray source preserving the photon flux. An increase of the source-object distance $R_0$ and the object detector distance $R_1$ has a beneficial effect on the phase-enhancement. For a non-coherent source such as bremsstrahlung or betatron it is still possible to define what is called lateral coherence $L_c$:

$$L_c = \frac{\lambda R_0}{s} \quad (1)$$

where $s$ is the source size. The lateral coherence defines the maximum distance between two points on the wave front which are able to interfere and to concur to the enhancement. $L_c$ increases with increasing the distance $R_0$, while keeping the source size as small as possible. Also, the x-ray energy plays a role, because higher energy leads to a smaller value $L_c$. However, because of the competition between absorption and phase-shift, the constraint on the wavelength for the lateral coherence can be compensated with the distance $R_0$, supporting the contribution of the phase against the absorption. The best set-up for XPCI has to keep into account all the x-ray source parameters in order to maximize the phase-enhancement against the absorption.

For discussed PHELIX parameters, 3D PIC simulations predict an ultra-high photon fluence (see Table 1) that remains at the level of $10^8/cm^2$ at 120 cm distance from the target (Fig. 3b). XPCI-simulations were made to forecast the imaging performance using DLA-based betatron radiation predicted by simulations in Section III. The code was designed to calculate x-ray absorption and phase contrast, taking into account the x-ray spectrum, source size and spatial intensity distribution and solving the Kirchoff-Fresnel equation using the Fresnel approximation[38]. The code works in cylindrical geometry and it takes into account along with the x-ray source parameters, the detector resolution and the density map associated to the object.

For simulations of the betatron based XPCI-experiment on the PHELIX-facility, a sphere of Polystyrene (CH) with a density $\rho = 1.046$ g/cm$^3$ and radius $r = 100$ μm was chosen as a test target. Plastic spheres are commonly used for studies of e.g. Rayleigh-Taylor instabilities[39] or strong shocks[40]. Plastic is quite transparent to hard radiation and this is a key point if one wants to underline the contribution of the phase-contrast against the absorption. The betatron source parameters such as x-ray spectrum, source size and opening angle were taken from the 3D PIC simulations presented in Section III. For the total distance $R_0 + R_1 = 120$ cm, where $R_0$ is a distance between the betatron source and the object and $R_1$ is between the object and the detector, a set of simulations varying the distance $R_0$ from 5 cm to 50 cm was performed at first. An estimated x-ray photon fluence reaches 100 ph /10 μm$^2$ for the total distance $R_1 + R_0 = 120$ cm. A large distance between the betatron source and the detector, which can be realized due to ultra-high photon fluence, allows for a higher spatial coherence, a higher image magnification and as a consequence a better spatial resolution. Figure 4 shows the comparison between the line out along the sphere axis at all the distances considered. The contribution of the phase-enhancement on the edge of the sphere is visible. The edge intensity increases up to 15 cm and then starts to decrease again. This initial set of simulations was performed assuming a single photon energy equal to 5 keV.

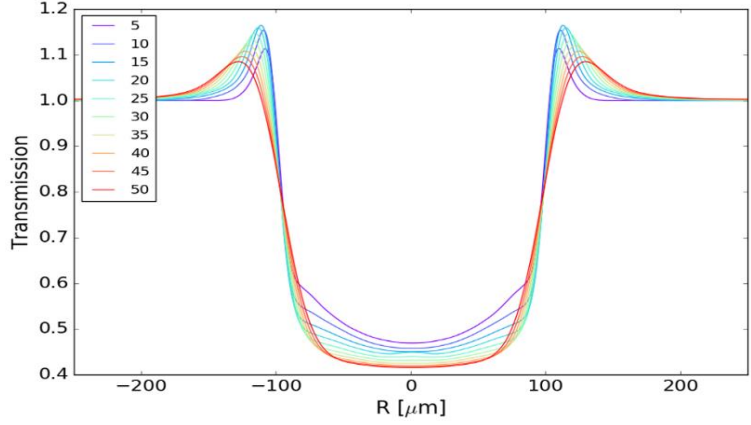

**FIG. 4.** Comparison between different distances from $R_0 = 5$ to $R_0 = 50$ cm. The total distance $R_0 + R_1$ is equal to 120 cm. The highest phase-enhancement was observed at $R_0 = 15$ cm.

A lateral coherence at the critical energy of 5 keV reaches Lc = 4 µm under the condition Ro = 15 cm and a source size s = 10 µm. For this source size, the spatial resolution with the optimized setup geometry and magnification M = 7 will be determined by the resolution of 100 µm of the imaging plate used in the simulation as a detector. For comparison, the lateral coherence reported in the case of LWFA[1] with $R_o$ = 1.1 m and a measured sources size of 1.7 µm was 9.6 µm at 10 keV photon energy, similar to what can be obtained with a synchrotron machine with an object located tens of meters from the x-ray source. In this experiment, the spatial resolution, determined using a setup with a magnification of 2.7 and an X-ray CCD camera with a pixel size of 20 µm, reached 7.4 µm.

Figure 5 presents a simulated phase-contrast image of the sphere where the whole betatron emission up to 40 keV (see spectrum in Fig. 3) and the sensitivity curve of the imaging plate FUJI BAS-TR were taken into account. The result is shown in Fig. 5(a), and the line out along the sphere axis is shown in Fig. 5(b). The effect of the phase-enhancement is clearly visible on the sphere border where the density has the vacuum-target discontinuity.

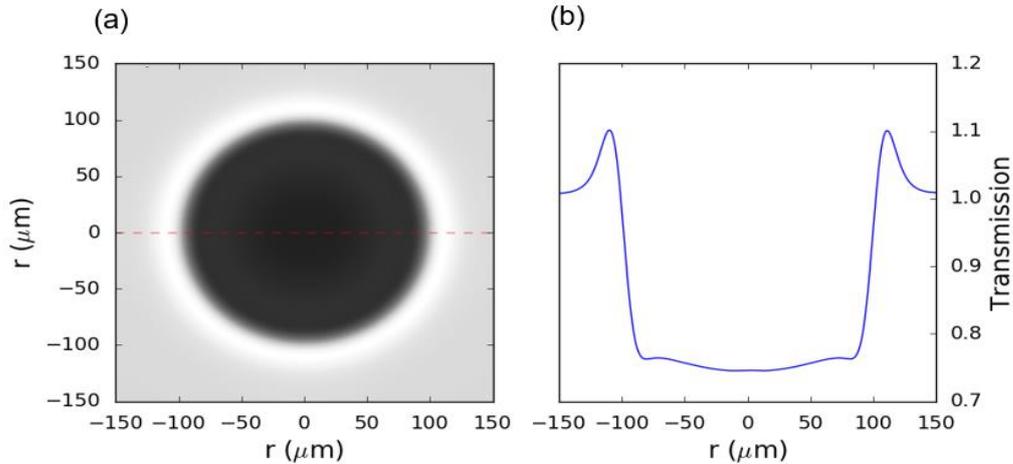

FIG. 5. (a) Simulated XPCI radiograph (b) the line out along the sphere axis (red dashed line). The presence of the phase-enhancement is clearly visible around the sphere. We considered the betatron emission up to 40 keV and the sensitivity curve of the IP-TR. For the x-ray source size, the worst case was assumed (initial laser spot size of 15 µm).

For comparing XPCI with a simple absorption radiography, a sphere with a density ten times lower than the solid CH sphere was irradiated with the simulated betatron spectrum (Fig. 3a) in the optimized set-up geometry. The resulting absorption image is shown in Fig. 6(a) while the XPCI in Fig. 6(b).

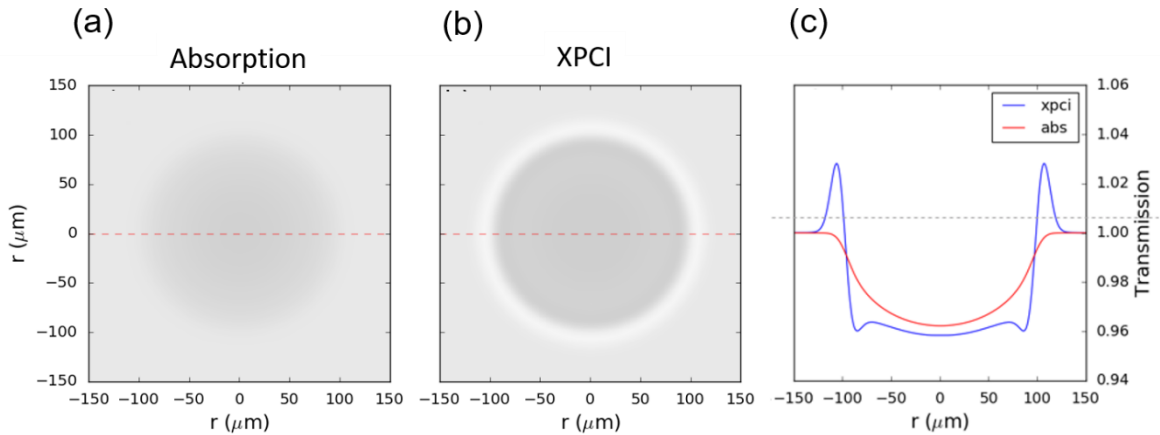

FIG. 6. Comparison between absorption (a) and XPCI (b). In (c) the line out across the cylinder axis shows the contribution of the phase-enhancement which makes the sphere otherwise too weak to detect in standard noisy experimental conditions.

For the considered sphere areal density ($\rho x \leq 100$ mg/cm$^3 \times 200$ µm) the absorption of photons with energies between 4 and 10 keV is very low (0.04% - 0.7%). Therefore, for a pure absorption radiography, it is clearly difficult to detect the sphere, while in the XPCI image the presence of the sphere is well visible due to the phase-enhancement. Also from the line out comparison shown in Fig. 6(c) we can observe the increasing contrast due to the phase contribution.

The comparison between absorption radiography and XPCI shown in Fig. 6 suggests that even using high energy x-ray radiation/low areal density, it is still possible to detect the object, while the intensity contrast in the absorption image makes absorption radiography ineffective. In presence of strong density gradients and density interfaces in the object, the contribution to the phase enhancement is weakly depending on the energy of the x-ray radiation. Simulations show, that the lower volume density of the object with a size of 100-200 µm is 10 mg/cm$^3$. This important feature makes the constraints on the x-ray source property less restricted compared to the standard absorption radiography.

## VI. CONLUSION

In this work, the feasibility of a DLA-based betatron source that promises an ultra-high number of x-ray photons is demonstrated. Numerical simulations of the intense betatron radiation that occurs in the DLA process and its radiographic applications are discussed for current parameters of the PHELIX-laser.

The simulations are based on the experimentally confirmed scheme of electron acceleration in NCD plasma in the interaction of a sub-picosecond laser pulse (≈0,7 ps) at a moderate relativistic laser intensity of ≈$2\times10^{19}$ W/cm$^2$ with sub-mm long pre-ionized polymer foams[22]. The use of low density foams has opened the door to the practical implementation of DLA in NCD plasma, requiring neither ultra-high laser intensity nor high laser contrast and has demonstrated its high robustness.

3D PIC-simulations of the interaction of 0.7 ps, $2\times10^{19}$ W/cm$^2$ laser pulse with long-scale NCD-plasma results in the betatron radiation with the ultra-high photon number of $7\times10^{11}$ per shot in the 1-30 keV range obtained for 20 J focused laser energy. A brilliance of the betatron source reached $\geq 6\times10^{19}$ ph/s/mm$^2$/mrad$^2$/0.1% BW at the critical energy of 5 keV. It is expected that increasing the laser energy keeping the laser pulse duration and focal spot size constant will lead to a grow in the energy and number of DLA electrons and to a corresponding increase of the betatron critical energy and x-ray photon number. This approach, which promises well-directed ultra-high fluence x-ray beams delivered on a picosecond time scale, will significantly enhance the radiographic capabilities of kJ PW-class laser systems operating in HED research.

ACKNOWLEDGMENTS

The experimental results presented here are based on the experiment P176, which was performed at the PHELIX facility at the GSI Helmholtzzentrum fuer Schwerionenforschung, Darmstadt (Germany) in the frame of FAIR Phase-0. The experimental group is very thankful for the support provided by the PHELIX-laser team at GSI-Darmstadt. This work was also supported by DFG (project PU 213/9), EPSRC grant EP/P026796/1. The authors gratefully acknowledge the Gauss Centre for Supercomputing e.V. for funding this project by providing computing time on the GCS Supercomputer JUWELS at Jülich Supercomputing Centre (JSC). X.F.S. gratefully acknowledges support by the Alexander von Humboldt Foundation. The research of V.S.P. and N.E.A. was supported by The Ministry of Science and Higher Education of the Russian Federation (Agreement with Joint Institute for High Temperatures RAS No 075-15-2020-785 dated September 23, 2020).